\makeatletter
\def\input@path{{latex_support/}}
\makeatother
\documentclass{JFM-FLM_Au}

\usepackage{etoolbox}
\makeatletter
\patchcmd{\@maketitle}{(Received xx; revised xx; accepted xx)\hfill}{}{}{}
\makeatother

\lefttitle{H.Y. Wang, H.Y. Zhu, X.Q. He and Y.L. Xiong}
\righttitle{Journal of Fluid Mechanics}

\title{Transition from a weak-elastic to an elastic-activated regime in two-dimensional polymeric turbulence}

\author{Haoyu Wang\aff{1,2}, Hangyu Zhu\aff{1,2}, Xiaoqiu He\aff{1,2} \and Yongliang Xiong\aff{1,2}}

\affiliation{\aff{1}Department of Mechanics, Huazhong University of Science and Technology, Wuhan 430074, China
\aff{2}Hubei Key Laboratory of Engineering Structural Analysis and Safety Assessment, Wuhan 430074, China}

\corresau{Yongliang Xiong, \email{xylcfd@hust.edu.cn}}

\begin{document}
\maketitle
\thispagestyle{plain}
\pagestyle{plain}

\begin{abstract}
We conduct direct numerical simulations to investigate the effects of dilute
polymers on flow structure, statistical properties and energy transfer in
2D turbulence. As the Weissenberg number \(Wi\) increases, the
flow undergoes a sharp but continuous crossover between a weak-elastic regime
and an elastic-activated regime. In the weak-elastic regime, the flow remains
close to its Newtonian counterpart and polymer stresses have weak dynamical
effects. In the elastic-activated regime, elastic stresses attenuate vortical
motions and promote shear-dominated filaments. Scale-by-scale energy analysis
shows that polymer addition weakens both the inverse kinetic-energy cascade and
the forward enstrophy cascade in the two regimes, whereas the kinetic--elastic
exchange and the inter-scale redistribution of elastic energy differ between
them. Kinetic-to-elastic conversion dominates over the analysed range of
scales in the weak-elastic regime, whereas a partial transfer from elastic to
kinetic energy emerges at small scales in the elastic-activated regime. The
inter-scale transfer of elastic energy is dominated by inverse transfer over
the large-scale range in the weak-elastic regime but by forward transfer over
the small-scale range in the elastic-activated regime. These results link the
crossover to the onset of strong elastic feedback from the polymers and the
associated reorganisation of the scale-by-scale transfer pathways.
Downstream-decay statistics also show the same qualitative two-regime
behaviour.

\end{abstract}



\section{Introduction}
\label{sec:headings}

Trace amounts of polymer can markedly alter flow behaviour. Since
Toms's seminal report \citep{Toms1949}, a central question has been how polymer
feedback modifies turbulent energy-transfer pathways. In conventional
three-dimensional turbulence, kinetic energy is transferred towards small
  dissipative scales by the nonlinear cascade \citep{Pope2000}, while polymer
stress introduces an additional kinetic--elastic exchange channel that can
redistribute this transfer
\citep{ANGELIS_CASCIOLA_BENZI_PIVA_2005,CasciolaDeAngelis2007,Xi2013,Zhang2021sciadv,Rosti2023,Chiarini_Singh_Rosti_2025}.

Two-dimensional turbulence provides a particularly rich setting for studying
polymer-mediated energy transfer. Although vortex stretching is absent, the
simultaneous conservation of kinetic energy and enstrophy gives rise to the
classical Kraichnan--Leith--Batchelor (KLB) dual-cascade phenomenology, with
inverse transfer of kinetic energy towards larger scales and forward transfer
of enstrophy towards smaller scales
\citep{Kraichnan1967,Leith1968,Batchelor1969}. The interaction between polymer
feedback and these two cascade processes therefore involves both the
modulation of the classical kinetic-energy and enstrophy transfers and the
exchange of energy between the flow and the polymer degrees of freedom.

These questions have attracted considerable experimental and numerical
interest. Flowing soap films provide a classical platform for studying 2D
turbulence and the effects of polymer additives, which have been shown to
suppress large-scale fluctuations and inverse kinetic-energy transfer while
modifying small-scale structures, vortex dynamics and turbulent transport
\citep{AmaroucheneKellay2002,Kellay2004,Jun2006,Hidema2018,HIDEMA2020104385,Fukushima2024,Fukushima2026}.
Numerical studies provided complementary evidence for changes in velocity fluctuations,
Lagrangian chaos, flow structures and the two classical transfer processes
\citep{Boffetta2003,Xiong_2011,Gupta2015}. In decaying 2D
elastoinertial turbulence, \citet{Gillissen2019} varied the polymer
concentration and distinguished a weakly coupled regime, a strongly coupled
regime and a laminar regime, in addition to the Newtonian reference.

Three-dimensional studies provide a useful, but not identical, reference for
such regime changes. Polymer activation is commonly associated with a
coil--stretch response governed by competition between polymer relaxation and
local flow time scales \citep{WatanabeGotoh2010,Rehman2022}, while elastic
ranges and non-monotonic polymer contributions have also been reported in
scale-by-scale analyses
\citep{Rosti2023,Chiarini_Singh_Rosti_2025}. The corresponding organisation of
the 2D dual cascade and the kinetic--elastic energy pathways
remains less clear.

Here, direct numerical simulations of a soap-film-like Oldroyd--B flow combine
region-resolved topology and decay statistics with local filter-space analysis,
spectra of the conformation square root and a kinetic Lin diagnostic. The
filter-space analysis follows established 2D transfer studies
\citep{Rivera_Daniel_Chen_Ecke_2003,LiaoOuellette2013}. A recent experimental
study of three-dimensional polymer turbulence also applied filter-space
analysis \citep{Wang2026JFM}. The square-root spectra and kinetic Lin diagnostic
follow the spectral framework of \citet{Nguyen2016}. Together, these measures
resolve kinetic, enstrophy and polymer-reservoir transfer alongside
shell-local kinetic--polymer conversion.

\section{Model description and numerical setup}
\label{sec:model}

The governing equations are the incompressible Navier--Stokes equations coupled
to the Oldroyd--B constitutive model
\citep{Oldroyd1950,Birdv2,Larson1988}:
\begin{subequations}\label{eq:gov}
\begin{gather}
\nabla\cdot\boldsymbol{u}=0, \label{eq:gov_cont}\\
\rho\left[\partial_t \boldsymbol{u} + \nabla\cdot(\boldsymbol{u}\otimes\boldsymbol{u})\right]
= -\nabla p + \nabla\cdot\left(2\mu_{\mathrm{s}} \boldsymbol{D}\right) + \nabla\cdot\boldsymbol{\sigma}, \label{eq:gov_mom}\\
\frac{\mathrm{D}\boldsymbol{C}}{\mathrm{D}t} - \boldsymbol{C}\cdot\nabla\boldsymbol{u} - (\nabla\boldsymbol{u})^{\mathrm{T}}\cdot\boldsymbol{C}
 = -\frac{\boldsymbol{C}-\boldsymbol{I}}{\lambda}, \label{eq:gov_C}
\end{gather}
\end{subequations}
Here \(\boldsymbol{u}\), \(p\) and \(\rho\) denote velocity, pressure and
density, respectively;
\(\boldsymbol{D}=[\nabla\boldsymbol{u}+(\nabla\boldsymbol{u})^{\mathrm{T}}]/2\),
and \(\mathrm{D}/\mathrm{D}t=\partial_t+\boldsymbol{u}\cdot\nabla\) are the
rate-of-strain tensor and material derivative, respectively. The dimensionless
conformation tensor \(\boldsymbol{C}\) describes the polymer configuration,
\(\boldsymbol{I}\) is the identity tensor, and
\(\boldsymbol{\sigma}=\mu_{\mathrm{p}}/\lambda(\boldsymbol{C}-\boldsymbol{I})\)
is the polymer stress. The solvent and polymer contributions to the zero-shear
viscosity are \(\mu_{\mathrm{s}}\) and \(\mu_{\mathrm{p}}\), respectively, so
that \(\mu_0=\mu_{\mathrm{s}}+\mu_{\mathrm{p}}\) and
\(\beta=\mu_{\mathrm{s}}/\mu_0\); \(\lambda\) is the polymer relaxation time.

Equations~\eqref{eq:gov} are integrated using the adaptive Cartesian solver
Basilisk and its viscoelastic implementation
\citep{LopezHerreraPopinetCastrejonPita2019}. The log-conformation
representation \(\boldsymbol{\Psi}=\log\boldsymbol{C}\) is used to alleviate the
high-Weissenberg-number problem
\citep{FattalKupferman2004,FattalKupferman2005}; no artificial stress diffusion
is added.

\begin{figure}
  \centerline{\includegraphics[width=0.85\textwidth]{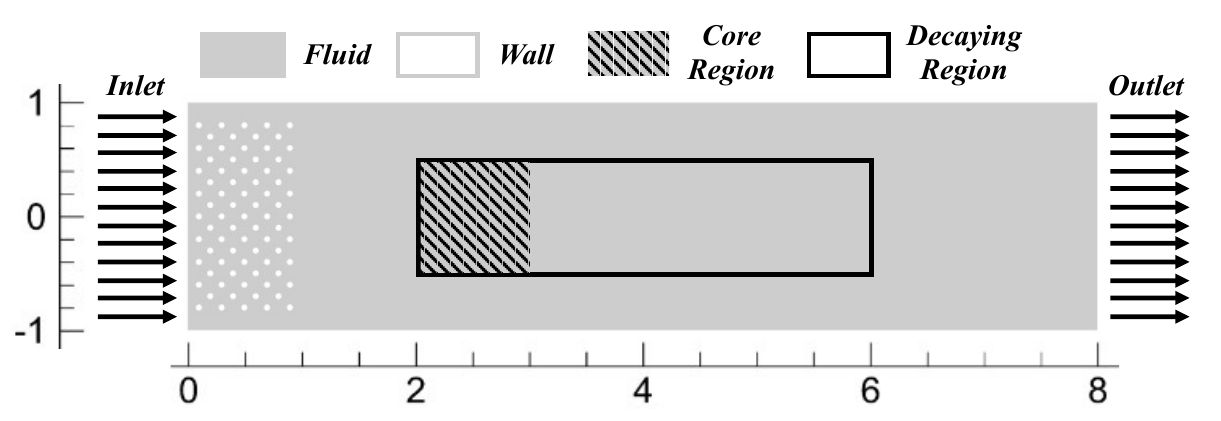}}
  \caption{Schematic of the soap-film-like configuration. An upstream cylinder
  array provides finite-amplitude perturbations, and statistics are collected
  downstream.}
\label{fig:geo}
\end{figure}

The computational configuration, shown in figure~\ref{fig:geo}, comprises a
soap-film-like flow perturbed by an upstream cylinder array. Unless stated
otherwise, statistics are sampled in a wall-remote \(1\times1\) core window
located one unit downstream of the array; a subsequent \(4\times1\) region is
used to quantify spatial distributions and downstream decay. With the channel
half-height \(H\) and bulk inflow velocity \(U_B\) as the reference length and
velocity, respectively, we set \(H=U_B=\rho=1\) and
\(\mu_0=3.3333\times10^{-4}\). The resulting bulk Reynolds number is
\(\Rey_B=U_BH/(\mu_0/\rho)=3000\), the nominal Weissenberg number is
\(Wi=\lambda U_B/H\), and the cylinder diameter is \(D=0.05H\). The principal
calculations employ a \(4096\times1024\) quadtree base grid with two additional
levels of near-wall refinement. The time step does not exceed
\(10^{-4}H/U_B\), and the Courant number remains below 0.1.

Resolution at the dissipative scales is quantified by
\(\eta=(\nu_{\mathrm{s}}^3/\varepsilon_{\mathrm{s}})^{1/4}\), where
\(\nu_{\mathrm{s}}=\mu_{\mathrm{s}}/\rho\) and
\(\varepsilon_{\mathrm{s}}\) is the solvent-viscous dissipation rate. The
associated time scale is
\(\tau_\eta=(\nu_{\mathrm{s}}/\varepsilon_{\mathrm{s}})^{1/2}\), from which
\(\Rey_\eta=u'_{\mathrm{rms}}\eta/\nu_{\mathrm{s}}\) and
\(Wi_\eta=\lambda/\tau_\eta\); \(u'_{\mathrm{rms}}\) denotes the core-region
r.m.s. velocity fluctuation. The base-grid spacing
\(\Delta_{\max}=0.00195\) is approximately \(\eta/3\). Repeating the principal
statistical, spectral and flux diagnostics for an \(8192\times2048\) Newtonian
case gave results consistent with the base-grid calculation. Statistics were
accumulated for approximately \(80D/U_B\) after transients; relative
differences between two equal subintervals remained below 1.5\% for the
tabulated quantities. Repeating the statistical analysis after shifting the
core window upstream and downstream yielded qualitatively consistent trends.
Table~\ref{tab:cases} summarises the simulated cases.

\begin{table}
  \begin{center}
  {\setlength{\tabcolsep}{5.5pt}
  \begin{tabular}{cccccccc}
      Case & Base grid & $Wi$ & $\lambda$ & $\beta$ & $\eta$ & $\Rey_{\eta}$ & $Wi_{\eta}$ \\[3pt]
      1  & \(4096\times1024\) & --    & --    & 1   & 0.0052  & 8.113 & -- \\
      2  & \(8192\times2048\) & --    & --    & 1   & 0.0053  & 7.973 & -- \\
      3  & \(4096\times1024\) & 0.001 & 0.001 & 0.9 & 0.0051  & 7.867 & 0.0116 \\
      4  & \(4096\times1024\) & 0.01  & 0.01  & 0.9 & 0.0052  & 7.816 & 0.1098 \\
      5  & \(4096\times1024\) & 0.05  & 0.05  & 0.9 & 0.0062  & 6.111 & 0.3840 \\
      6  & \(4096\times1024\) & 0.1   & 0.1   & 0.9 & 0.0084  & 3.757 & 0.4243 \\
      7  & \(4096\times1024\) & 0.3   & 0.3   & 0.9 & 0.0119  & 2.576 & 0.6391 \\
      8  & \(4096\times1024\) & 0.4   & 0.4   & 0.9 & 0.0118  & 2.270 & 0.8654 \\
      9  & \(4096\times1024\) & 0.5   & 0.5   & 0.9 & 0.0112  & 2.139 & 1.1967 \\
      10 & \(4096\times1024\) & 0.8   & 0.8   & 0.9 & 0.0101  & 1.962 & 2.3388 \\
      11 & \(4096\times1024\) & 1.5   & 1.5   & 0.9 & 0.0095  & 1.918 & 4.9684 \\
  \end{tabular}}
  \caption{Simulation cases, base-grid resolutions and corresponding dissipative-scale parameters. Case 2 is a high-resolution Newtonian case used for grid-resolution checking.}
  \label{tab:cases}
  \end{center}
\end{table}

\section{Statistical properties and the transition}
\label{sec:statistics}

\begin{figure}
  \centerline{\includegraphics[width=1\textwidth]{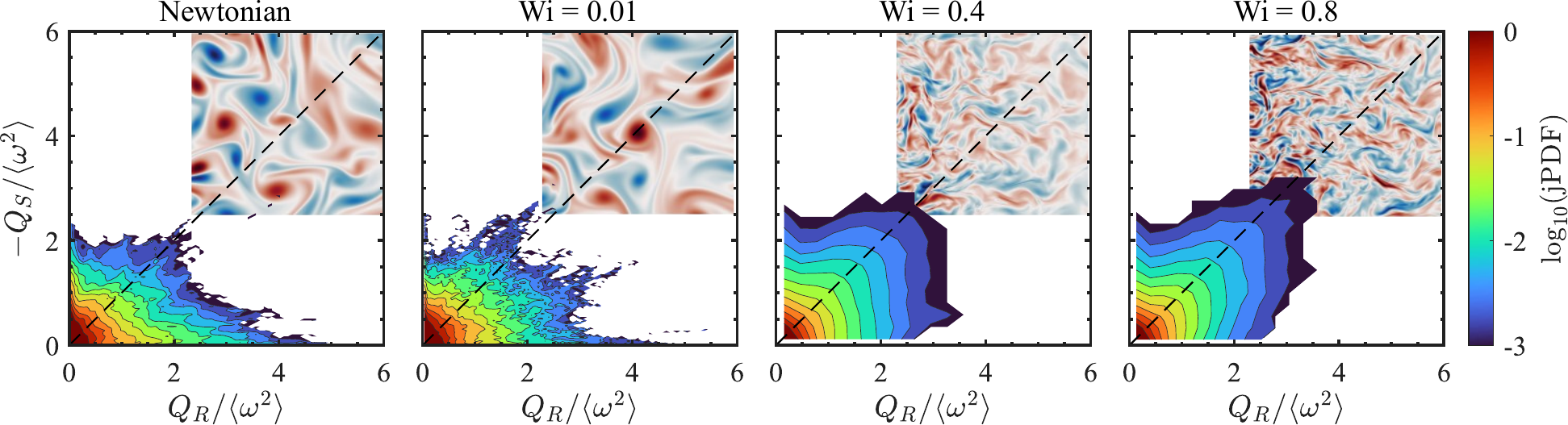}}
  \caption{Joint PDFs of the normalised invariants \(Q_S\) and \(Q_R\), with
  corresponding instantaneous vorticity fields shown as insets. Each inset uses
  an independent colour scale to resolve the corresponding flow morphology.}
\label{fig:jPDF}
\end{figure}

To quantify polymer-induced changes in local flow structure, we examine the
joint probability density function (jPDF) of the strain- and rotation-related
velocity-gradient invariants \(Q_S\) and \(Q_R\):
\begin{equation}
Q_S=-\frac{1}{2}\mathrm{tr}(\boldsymbol{D}^2),
\qquad
Q_R=\frac{\omega^2}{4}.
\label{eq:QSQR_invariants}
\end{equation}
Figure~\ref{fig:jPDF} shows the resulting jPDFs in the \((Q_R,-Q_S)\) plane,
with \(Q_R\) and \(-Q_S\) normalised by \(\langle\omega^2\rangle\), together
with representative instantaneous vorticity fields. For the Newtonian and
lower-\(Wi\) cases, the distributions extend preferentially towards the
\(Q_R\) axis, consistent with the presence of large coherent vortices. As
\(Wi\) increases, the \(Q_R\) tail contracts while the extent in \(-Q_S\)
changes little, bringing the distribution closer to the diagonal
\(Q_R=-Q_S\), where rotation and strain make comparable local contributions.
The accompanying replacement of large vortices by elongated filaments is
consistent with a shear-layer-like organisation and resembles the vortex-sheet
structures reported in three-dimensional bulk polymeric turbulence
\citep{Zhang2025,Wang2025VGT}.

\begin{figure}
  \centerline{\includegraphics[width=1.0\textwidth]{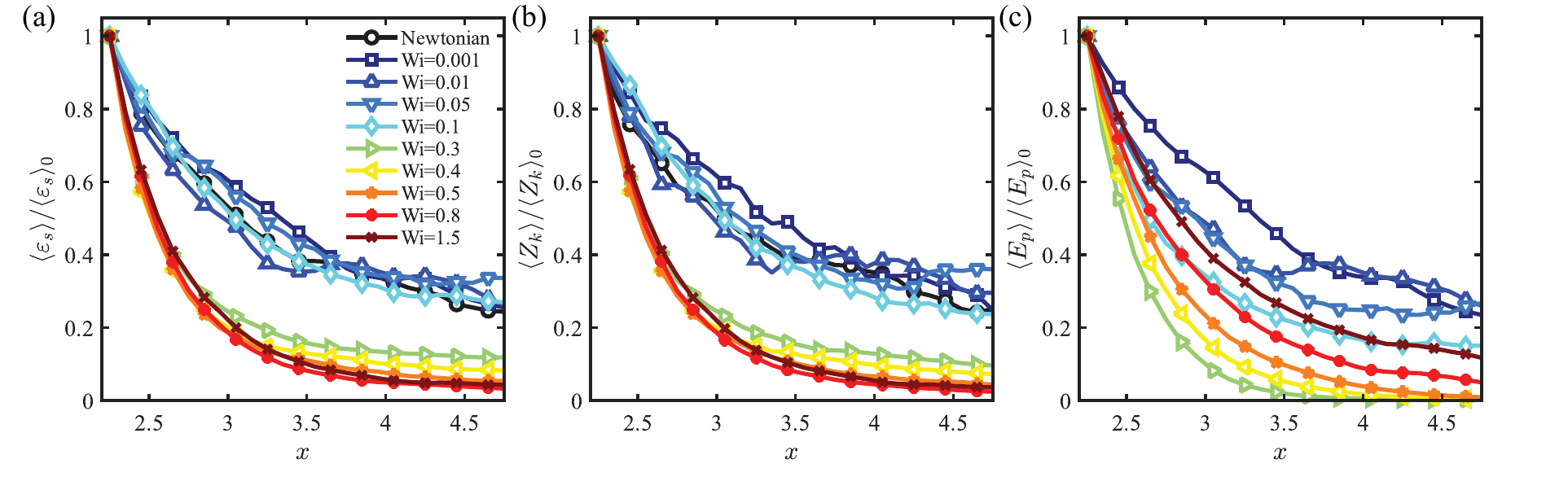}}
  \caption{Streamwise decay of normalised statistical quantities in the
  downstream region. (a) Turbulent dissipation rate
  \(\langle\varepsilon_{\mathrm{s}}\rangle/\langle\varepsilon_{\mathrm{s}}\rangle_0\).
  (b) Enstrophy \(\langle Z_K\rangle/\langle Z_K\rangle_0\).
  (c) Elastic energy \(\langle E_P\rangle/\langle E_P\rangle_0\). For each
  case, the subscript 0 denotes its first streamwise sampling bin.}
\label{fig:decay}
\end{figure}

In contrast to canonical homogeneous turbulence, the present
spatially developing configuration provides a complementary view through its
pronounced streamwise decay. Figure~\ref{fig:decay} quantifies this evolution
using the normalised turbulent dissipation rate, enstrophy and
elastic energy. The latter is constructed from the symmetric square root
\(B_{ij}=(C^{1/2})_{ij}\) following \citet{Balci2011}. The turbulent dissipation
rate and enstrophy show similar downstream behaviour
(figure~\ref{fig:decay}a,b). Both remain close to the Newtonian case at low
\(Wi\), but decay substantially faster as \(Wi\) approaches the crossover range
\(0.1\)--0.3. The elastic energy follows a distinct trend
(figure~\ref{fig:decay}c): its normalised decay steepens within this range and
reaches its largest sampled rate near \(Wi=0.3\). At higher \(Wi\), this trend
reverses and the normalised decay becomes slower. Thus, the fluid and polymer
measures respond differently to \(Wi\), but reorganise over the same interval.

\begin{figure}
  \centerline{\includegraphics[width=1.0\textwidth]{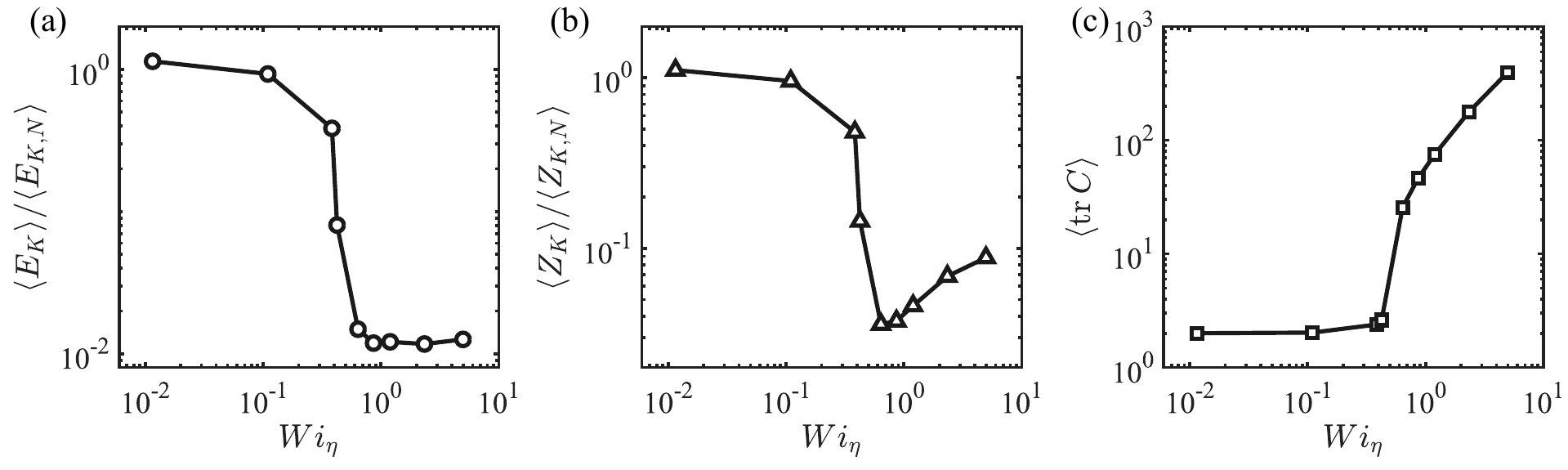}}
  \caption{Core-region indicators of the transition as functions of
  \(Wi_{\eta}\). (a) Kinetic energy normalised by the Newtonian
  value. (b) Enstrophy normalised by the Newtonian value.
  (c) Mean trace of the conformation tensor.}
\label{fig:transition1}
\end{figure}

We use \(Wi_{\eta}\) to order the core-region statistics in
figure~\ref{fig:transition1}. The normalised kinetic energy and enstrophy remain
close to their Newtonian values at low \(Wi_{\eta}\), decrease sharply over an
intermediate range and remain low thereafter, with a weak non-monotonic
recovery. Over the same range,
\(\langle\mathrm{tr}\,\boldsymbol{C}\rangle\) increases rapidly, indicating strong
polymer stretching. These concurrent changes in the flow and polymer measures
define an operational transition range over \(Wi=0.1\)--0.3
(\(Wi_{\eta}\simeq0.42\)--0.64). We refer to the lower-\(Wi\) response as weakly
elastic and the higher-\(Wi\) response as elastic-activated. The topology,
downstream decay and alternative core sampling windows exhibit concurrent
changes over this interval. To further elucidate the mechanisms underlying this
transition, we next examine the energy transfers scale by scale.

\section{Scale-by-scale analysis}
\label{sec:scale}

We define the quadratic polymer reservoir using the unique symmetric
positive-definite square root \(B_{ij}=(C^{1/2})_{ij}\), introduced as a
conformation-tensor formulation by \citet{Balci2011} and applied to turbulent
spectra by \citet{Nguyen2016}. The corresponding quadratic polymer energy is
\(P=K_p(B_{ij}B_{ij}-d)=K_p(C_{ii}-d)\), where
\(K_p=\mu_p/(2\rho\lambda)\) and \(d=2\).
Following the viscoelastic spectral formulations of
\citet{CasciolaDeAngelis2007} and \citet{Nguyen2016}, the core-region fields
are multiplied by a 2D Hann window. The shell spectra are
\begin{equation}
E_K(\kappa)=\frac12\sum_{\boldsymbol{k}\in\mathcal K_\kappa}
\left|\widehat u_i(\boldsymbol{k})\right|^2,\qquad
E_P(\kappa)=K_p\sum_{\boldsymbol{k}\in\mathcal K_\kappa}
\left|\widehat B_{ij}(\boldsymbol{k})\right|^2.
\label{eq:shell_spectra}
\end{equation}
The corresponding polymer-stress contribution to the kinetic-energy equation
is
\begin{equation}
S_{P\rightarrow K}(\kappa)=\frac{1}{\rho}
\sum_{\boldsymbol{k}\in\mathcal K_\kappa}\operatorname{Re}\!\left[
\widehat u_i^{\,*}(\boldsymbol{k})\,\mathrm{i}k_j
\widehat\sigma_{ij}(\boldsymbol{k})\right].
\label{eq:SPK_shell}
\end{equation}
Here \(\widehat{(\,\cdot\,)}\) denotes the Fourier transform and
\(\kappa=|\boldsymbol{k}|\).

\begin{figure}
  \centerline{\includegraphics[width=1.0\textwidth]{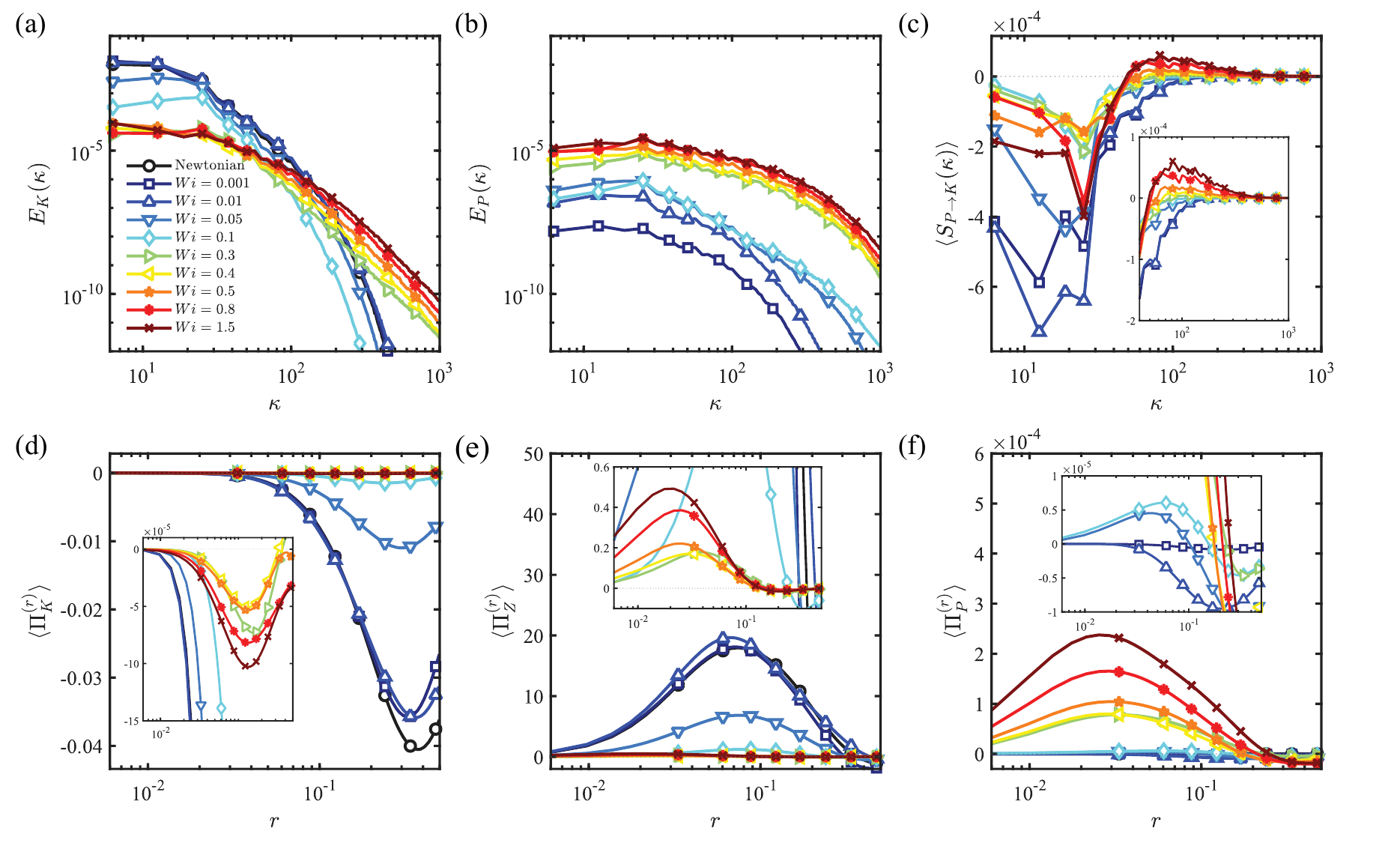}}
  \caption{Core-region spectra and transfer diagnostics. (a) 2D
  fluctuation kinetic-energy spectrum \(E_K(\kappa)\). (b)
  Elastic-energy spectrum \(E_P(\kappa)\). (c) Lin shell exchange
  \(S_{P\rightarrow K}(\kappa)\), for which positive values denote kinetic-energy
  gain by the shell at \(\kappa\). Core-region means of the Gaussian-filter
  transfers: (d) kinetic energy, \(\Pi_K^{(r)}\); (e) enstrophy,
  \(\Pi_Z^{(r)}\); and (f) polymer energy, \(\Pi_P^{(r)}\).
  Insets enlarge the indicated ranges.}
\label{fig:scale_transfer}
\end{figure}

Figure~\ref{fig:scale_transfer} a--c characterises the spectral distributions of
kinetic and elastic energy and the scale-dependent exchange between them. At
the two lowest non-zero \(Wi\), \(E_K\) remains close to the Newtonian spectrum
and \(E_P\) has low amplitude. With increasing elasticity, \(E_K\) is reduced,
particularly at low wavenumber. Within the elastic-activated regime,
however, its high-wavenumber range progressively recovers as \(Wi\) increases,
consistent with the 2D homogeneous results of \citet{Gupta2015}, while \(E_P\)
broadens and gains amplitude.
Positive \(S_{P\rightarrow K}\) denotes kinetic-energy gain by the shell at
\(\kappa\), whereas negative values denote extraction by polymer stress. The
Lin exchange remains negative at low wavenumber and develops a positive
lobe in the high-wavenumber range in the elastic-activated cases. This identifies a
shell-local return of polymer energy that coexists with low-wavenumber
kinetic-energy extraction and is consistent with the high-wavenumber recovery
of \(E_K\). A similar small-scale polymer-to-fluid contribution was obtained
from the 3D spectral budget of \citet{Nguyen2016}.

For the filter-space analysis, we extend established FST formulations for
2D turbulence \citep{Rivera_Daniel_Chen_Ecke_2003,LiaoOuellette2013}
and coarse-grain a field \(q\) as
\begin{equation}
q^{(r)}(\boldsymbol{x},t)
=\int_{\mathbb R^2}\mathcal G^{(r)}
(\boldsymbol{x}-\boldsymbol{x}')q(\boldsymbol{x}',t)\,
\mathrm{d}\boldsymbol{x}',
\label{eq:main_filter}
\end{equation}
where \(\mathcal G^{(r)}\) is a normalised 2D Gaussian kernel.
Applying this filter to the governing equations partitions the kinetic and
quadratic polymer energies into resolved and subfilter-scale (SFS) reservoirs.
The resulting local balances, derived in the supplementary material, are
\begin{subequations}\label{eq:main_four_budget}
\begin{align}
\partial_tK^{(r)}+\partial_{x_j}J_{K,j}^{(r)}
={}&-\Pi_K^{(r)}-\mathcal W_L^{(r)}
-\varepsilon_{\mathrm{s}}^{(r)}+F^{(r)},\\
\partial_tK_{\mathrm{SFS}}^{(r)}
+\partial_{x_j}J_{K,\mathrm{SFS},j}^{(r)}
={}&+\Pi_K^{(r)}-\mathcal W_S^{(r)}
-\varepsilon_{\mathrm{s},\mathrm{SFS}}^{(r)}+F_{\mathrm{SFS}}^{(r)},\\
\partial_tP^{(r)}+\partial_{x_j}J_{P,j}^{(r)}
={}&-\Pi_P^{(r)}
+\mathcal S_{B,G}^{(r)}-R^{(r)},\\
\partial_tP_{\mathrm{SFS}}^{(r)}
+\partial_{x_j}J_{P,\mathrm{SFS},j}^{(r)}
={}&+\Pi_P^{(r)}
+\left(\mathcal W_T^{(r)}-\mathcal S_{B,G}^{(r)}\right)
-R_{\mathrm{SFS}}^{(r)}.
\end{align}
\end{subequations}
Here \(K^{(r)}=u_i^{(r)}u_i^{(r)}/2\) and
\(P^{(r)}=K_p(B_{ij}^{(r)}B_{ij}^{(r)}-d)\) are the resolved kinetic and
quadratic polymer energies, respectively. The corresponding SFS energies are
\(K_{\mathrm{SFS}}^{(r)}
=([u_i u_i]^{(r)}-u_i^{(r)}u_i^{(r)})/2\) and
\(P_{\mathrm{SFS}}^{(r)}
=K_p([B_{ij}B_{ij}]^{(r)}-B_{ij}^{(r)}B_{ij}^{(r)})\).
The fluxes \(J_{q,j}\) collect physical-space transport. The
pairs \(\mp\Pi_K^{(r)}\) and \(\mp\Pi_P^{(r)}\) conservatively transfer energy
between the resolved and SFS kinetic and polymer reservoirs, respectively, with
\(\Pi_P^{(r)}=\Pi_{P,\mathrm{adv}}^{(r)}+\Pi_{\mathrm{ori}}^{(r)}\)
combining subfilter advection and square-root orientation compensation. The
signed terms \(\mathcal W_L^{(r)}\) and
\(\mathcal W_S^{(r)}\) are polymer-stress work in the resolved and
SFS kinetic equations; their polymer-side stretching entries are
\(\mathcal S_{B,G}^{(r)}\) and
\(\mathcal W_T^{(r)}-\mathcal S_{B,G}^{(r)}\), with
\(\mathcal W_T^{(r)}=[\sigma_{ij}\partial_{x_i}u_j/\rho]^{(r)}
=\mathcal W_L^{(r)}+\mathcal W_S^{(r)}\).
Here \(\varepsilon_{\mathrm{s}}\), \(F\) and \(R\) denote solvent dissipation,
external power and polymer relaxation. 
The kinetic and enstrophy transfers have the standard FST forms,
whereas the polymer transfer is obtained by filtering the square-root equation:
\begin{subequations}\label{eq:FST_transfers}
\begin{align}
\Pi_K^{(r)}&=-\tau^u_{ij}G_{ij}^{(r)},\\
\Pi_Z^{(r)}&=-\tau^\omega_j\partial_{x_j}\omega^{(r)},\\
\Pi_P^{(r)}&=-2K_p\tau^B_{kij}\partial_{x_k}B_{ij}^{(r)}
-2K_pB_{ij}^{(r)}[a_{i\ell}B_{\ell j}]^{(r)},
\end{align}
\end{subequations}
where \(\tau^u_{ij}=[u_i u_j]^{(r)}-u_i^{(r)}u_j^{(r)}\),
\(\tau^\omega_j=[u_j\omega]^{(r)}-u_j^{(r)}\omega^{(r)}\) and
\(\tau^B_{kij}=[u_kB_{ij}]^{(r)}-u_k^{(r)}B_{ij}^{(r)}\).
Here \(a_{ij}\) is the antisymmetric compensation field of the principal
symmetric square root. 
Under this sign convention, negative values denote transfer from smaller to
larger scales, whereas positive values denote transfer from larger to smaller
scales.

Figure~\ref{fig:scale_transfer}d--f shows the corresponding core-region
filter-space transfers. The dominant branches of \(\Pi_K^{(r)}\) in panel (d)
and \(\Pi_Z^{(r)}\) in panel (e) remain negative and positive, respectively,
corresponding to upscale kinetic-energy transfer and downscale enstrophy
transfer, as in the Newtonian flow. Relative to the Newtonian case, both
transfers are progressively attenuated as \(Wi\) increases within the weakly
elastic regime and decrease sharply through the transition. After elastic
activation, their amplitudes recover partially but remain well below the
Newtonian levels. By contrast, the polymer self-transfer
\(\Pi_P^{(r)}\) in panel (f) changes from a negative branch dominated over the
large-scale range in the weakly elastic regime to a positive branch dominated
over the small-scale range after elastic activation. 

\begin{figure}
  \centerline{\includegraphics[width=1.0\textwidth]{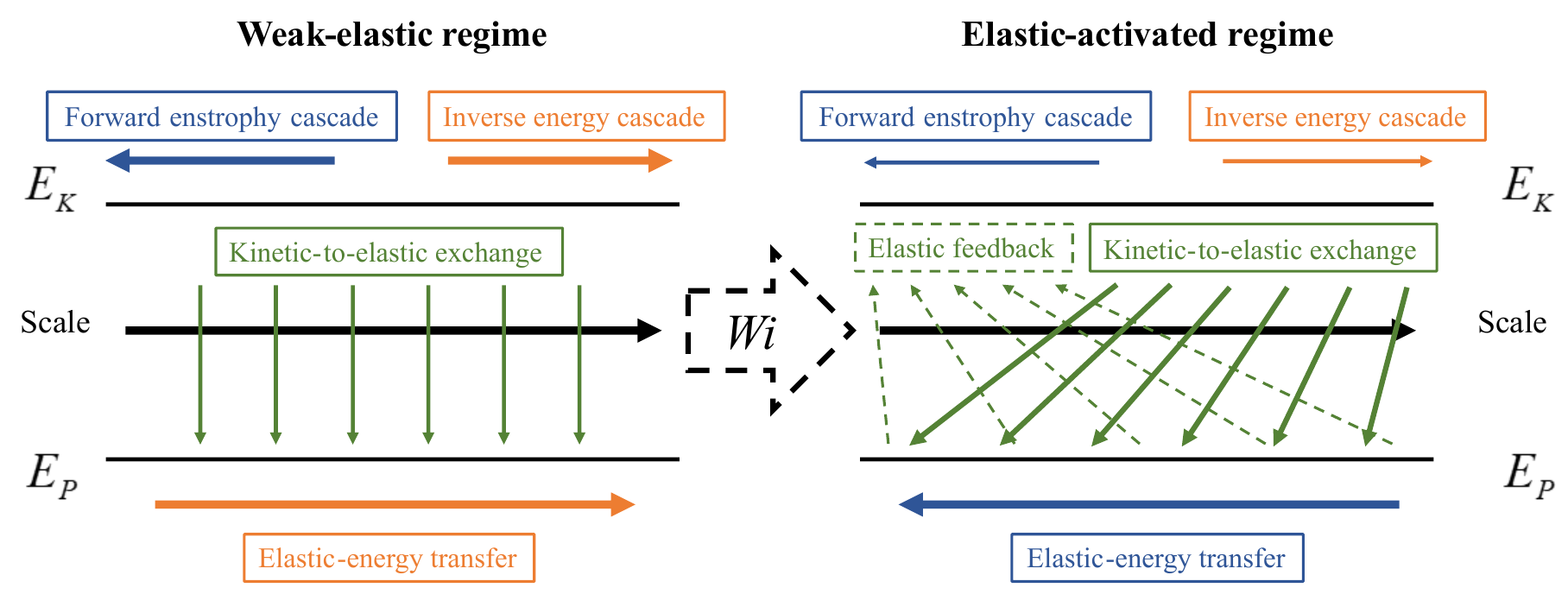}}
  \caption{Schematic of the diagnosed energy-transfer pathways between
  kinetic energy \(E_K\) and the elastic-energy reservoir \(E_P\).
  Orange and blue horizontal arrows denote inverse and forward inter-scale
  transfer, respectively. Solid and dashed green arrows denote
  kinetic-to-elastic conversion and elastic-to-kinetic feedback, respectively;
  the latter corresponds to the positive high-wavenumber contribution to
  \(S_{P\rightarrow K}\).}
\label{fig:mechanism}
\end{figure}

The Lin spectral budget and filter-space analysis together provide the 
energy-transfer picture summarised in Figure~\ref{fig:mechanism}. 
In the weakly elastic 
regime, the classical upscale kinetic-energy transfer and downscale 
enstrophy transfer remain dominant. Polymer stress extracts kinetic 
energy over the analysed wavenumber range, while polymer energy is 
transferred predominantly upscale. 
After elastic activation, the two classical transfers persist but are 
strongly attenuated. Polymer stress continues to extract kinetic energy 
at low wavenumbers, whereas a positive high-wavenumber Lin contribution 
reveals a shell-local polymer-to-kinetic return. 
Concurrently, the polymer self-transfer changes from a 
large-scale-dominated inverse branch to a small-scale-dominated forward 
branch, indicating a reorganisation towards downscale polymer-energy 
transfer.

\begin{figure}
  \centerline{\includegraphics[width=1.0\textwidth]{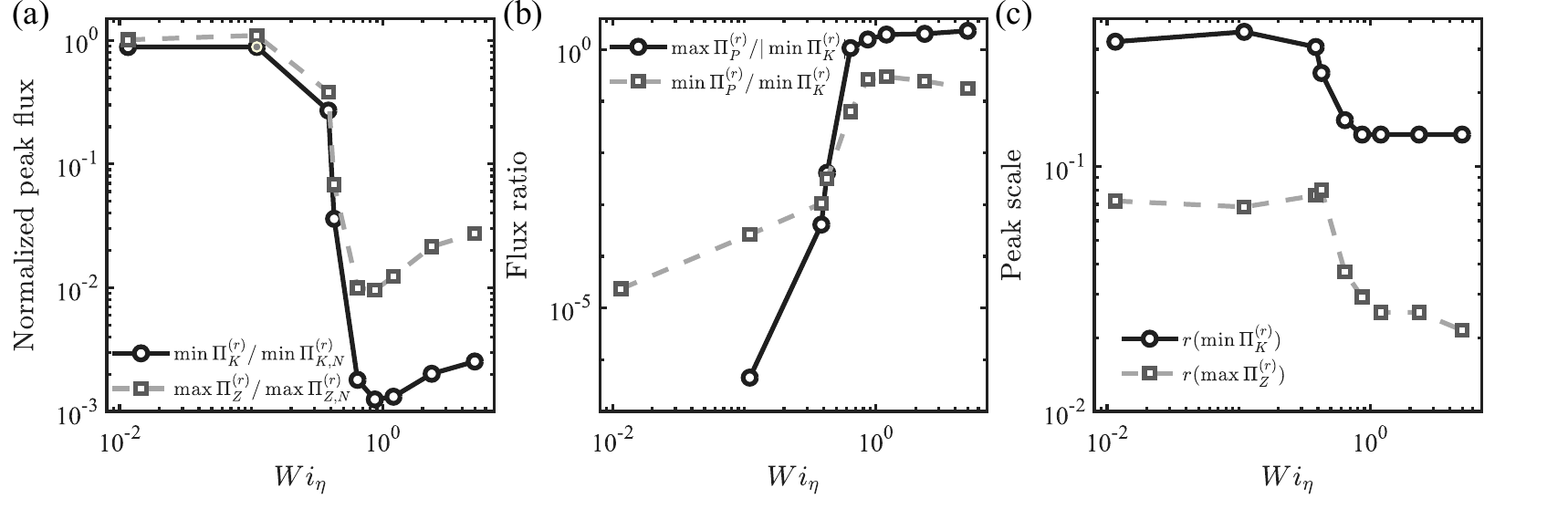}}
  \caption{Characteristic filter-space measures versus \(Wi_{\eta}\). (a)
  Magnitudes of the inverse kinetic-energy and forward enstrophy-transfer peaks,
  normalised by their Newtonian values. (b) Positive and negative extrema of
  the polymer-energy transfer, \(\max_r\Pi_P^{(r)}\) and
  \(-\min_r\Pi_P^{(r)}\), divided by \(-\min_r\Pi_K^{(r)}\) for each case.
  (c) Filter scales of the inverse kinetic-energy and forward enstrophy peaks.}
\label{fig:transition3}
\end{figure}

Figure~\ref{fig:transition3} quantifies the changes in the amplitudes and
characteristic scales of the filter-space transfers. In the weakly elastic
regime, the inverse kinetic-energy and forward enstrophy-transfer peaks remain
close to their Newtonian values. Across the transition, both decrease by
approximately two to three orders of magnitude and recover only weakly at higher
$Wi_\eta$. Meanwhile, $\Pi_P^{(r)}$ changes from a dominant negative extremum
to a dominant positive extremum, indicating a transition from inverse to
forward-dominated polymer-energy transfer and an increasing relative importance
of the polymer pathway. Meanwhile, the peak locations of the 
kinetic-energy and enstrophy transfers move towards smaller scale, 
indicating that the residual transfer activity becomes concentrated at 
progressively smaller scales.

\section{Conclusions}
\label{sec:conclusion}

Using direct numerical simulations, we find that
polymer feedback organises 2D turbulence into weakly elastic and
elastic-activated regimes connected by a finite transition range. Topology,
core-region statistics, downstream decay and scale-resolved transfers provide
consistent signatures of this organisation.

At low elasticity, polymer extension and the quadratic polymer reservoir remain
weak. The flow remains close to its Newtonian counterpart, and inverse
kinetic-energy and forward enstrophy transfer remain prominent. Once elasticity
is activated, polymer extension rises, vortical structures weaken and the
flow-side quantities decay faster downstream. Both classical transfers are
strongly attenuated. In addition, the Lin spectral diagnostic develops a
positive lobe at high wavenumbers, revealing shell-local polymer-to-kinetic
return alongside kinetic-energy extraction at low wavenumbers. The inter-scale
polymer-energy transfer also changes from a negative branch dominated over the
large-scale range in the weakly elastic regime to a positive branch dominated
over the small-scale range after elastic activation. These observations show
that elastic activation is accompanied by both attenuation of the classical
kinetic-energy and enstrophy transfers and reorganisation of kinetic--polymer
conversion and inter-scale polymer-energy transport.

The crossover occupies a finite interval rather than a single critical point.
By varying polymer concentration, \citet{Gillissen2019} distinguished weakly
coupled, strongly coupled and laminar regimes in decaying 2D elastoinertial
turbulence. The present relaxation-time-controlled system instead exhibits
weakly elastic and elastic-activated regimes. The two studies therefore both
demonstrate distinct dynamical responses to polymer feedback in 2D flows,
although their control parameters and regime classifications differ. The
present crossover should also not be identified directly with the
three-dimensional coil--stretch transition
\citep{WatanabeGotoh2010,Rehman2022}, because the present 2D system lacks vortex
stretching and retains dual-cascade dynamics.

\bibliographystyle{latex_support/jfm}
\bibliography{latex_support/jfm}

\end{document}


\maketitle

\section*{Derivation of the filtered energy equations}

This supplementary material derives the filtered kinetic- and polymer-energy
equations used in the main text. Starting from the Oldroyd--B equations and
the symmetric positive-definite square root of the conformation tensor, we
separate the kinetic and quadratic polymer energies into resolved and
subfilter-scale reservoirs and identify the associated inter-scale transfer,
kinetic--polymer exchange and relaxation terms. Einstein summation is used
over repeated Latin indices \(i,j,k,\ell=1,2\), and all energies are per unit
mass.

\subsection*{Filter and notation}

For any primitive field \(q\), a homogeneous, time-independent filter is
written as
\begin{equation}
q^{(r)}(x_1,x_2,t)
=\int_{\mathbb{R}^2}
\mathcal G^{(r)}(x_1-x'_1,x_2-x'_2)
q(x'_1,x'_2,t)\,\dd x'_1\,\dd x'_2,
\label{eq:supp_filter}
\end{equation}
and filtering commutes with temporal and spatial derivatives.  If the filter
acts on a composite expression \(Q\), its scope is shown explicitly by square
brackets; thus \([q_1q_2]^{(r)}\neq q_1^{(r)}q_2^{(r)}\) in general.
Spatial, core-region and temporal averages are
denoted by angle brackets with the corresponding subscript. A superscript
\((r)\) therefore identifies a filtered field, a filtered composite, or a
scale-dependent quantity constructed from filtered fields. Subfilter
reservoirs are additionally marked by the subscript \(\mathrm{SFS}\). The
numerical analysis uses the normalised two-dimensional Gaussian kernel
\begin{equation}
\mathcal G^{(r)}(\xi_1,\xi_2)
=\frac{1}{2\pi\sigma_r^2}
\exp\!\left(-\frac{\xi_j\xi_j}{2\sigma_r^2}\right),
\qquad
\sigma_r=\frac{r}{\sqrt{2}\pi}.
\label{eq:supp_gaussian_filter}
\end{equation}
The Gaussian convolution is truncated at $4\sigma_r$ and evaluated using a
halo of the same width. We define
\begin{equation}
G_{ij}=\partial_{x_i}u_j,
\qquad
D_{ij}=\frac{1}{2}(G_{ij}+G_{ji}),
\qquad
\partial_{x_i}u_i=0.
\label{eq:supp_gradient}
\end{equation}

\subsection*{Governing equations}

For constant material properties, the momentum and Oldroyd--B conformation
equations are
\begin{subequations}\label{eq:supp_governing}
\begin{align}
\partial_tu_i+u_j\partial_{x_j}u_i
={}&-\frac{1}{\rho}\partial_{x_i}p
+2\nu_{\mathrm{s}}\partial_{x_j}D_{ij}
+\frac{1}{\rho}\partial_{x_j}\sigma_{ij}+f_i,
\label{eq:supp_momentum}\\
\frac{\mathrm{D}C_{ij}}{\mathrm{D}t}
\equiv\partial_tC_{ij}+u_k\partial_{x_k}C_{ij}
={}&C_{ik}G_{kj}+G_{ki}C_{kj}
+\frac{1}{\lambda}(\delta_{ij}-C_{ij}),
\label{eq:supp_C}
\end{align}
\end{subequations}
with
\begin{equation}
\sigma_{ij}
=\frac{\mu_{\mathrm{p}}}{\lambda}(C_{ij}-\delta_{ij}),
\qquad
K_p=\frac{\mu_p}{2\rho\lambda}.
\label{eq:supp_stress_Kp}
\end{equation}

For a symmetric positive-definite conformation tensor, introduce its unique
symmetric positive-definite square root
\begin{equation}
C_{ij}=B_{ki}B_{kj}=B_{ik}B_{kj},
\qquad
B_{ij}=B_{ji},
\qquad
B_{ij}\xi_i\xi_j>0\quad\text{for}\quad\xi_i\xi_i>0.
\label{eq:supp_C_B}
\end{equation}
In two dimensions this factor can be evaluated without an eigendecomposition:
\begin{equation}
B_{ij}=\frac{C_{ij}+\sqrt{\det[C_{mn}]}\,\delta_{ij}}
{\sqrt{C_{kk}+2\sqrt{\det[C_{mn}]}}}.
\label{eq:supp_B_2d}
\end{equation}
Following the symmetric factorisation of \citet{Balci2011}, the exact
Oldroyd--B equation for \(B_{ij}\) is
\begin{equation}
\frac{\mathrm{D}B_{ij}}{\mathrm{D}t}
\equiv\partial_tB_{ij}+u_k\partial_{x_k}B_{ij}
=B_{i\ell}G_{\ell j}+a_{i\ell}B_{\ell j}
+\frac{1}{2\lambda}\left[(B^{-1})_{ij}-B_{ij}\right],
\label{eq:supp_B}
\end{equation}
where \(a_{ji}=-a_{ij}\) is fixed by requiring the
right-hand side of equation~\eqref{eq:supp_B} to remain symmetric:
\begin{equation}
a_{i\ell}B_{\ell j}+B_{i\ell}a_{\ell j}
=G_{\ell i}B_{\ell j}-B_{i\ell}G_{\ell j}.
\label{eq:supp_a_sylvester}
\end{equation}
In two dimensions,
\begin{equation}
a_{11}=a_{22}=0,
\qquad
a_{21}=-a_{12},
\qquad
a_{12}=\frac{
B_{12}G_{11}+B_{22}G_{21}-B_{11}G_{12}-B_{12}G_{22}}
{B_{11}+B_{22}}.
\label{eq:supp_a_2d}
\end{equation}

\subsection*{Kinetic and polymer energies}

Define
\begin{equation}
K=\frac{1}{2}u_i u_i,
\qquad
P=K_p(B_{ij}B_{ij}-d)=K_p(C_{ii}-d),
\qquad
\phi=\frac{\sigma_{ij}G_{ij}}{\rho}
=2K_pC_{ij}G_{ij}.
\label{eq:supp_K_P_phi}
\end{equation}
Here \(d=2\) for the present two-dimensional system.
Multiplying equation~\eqref{eq:supp_momentum} by \(u_i\) and applying the
product rule gives
\begin{equation}
\partial_tK+\partial_{x_j}J_{K,j}
=-\varepsilon_{\mathrm{s}}-\phi+F,
\label{eq:supp_K_unfiltered}
\end{equation}
where
\begin{align}
J_{K,j}={}&u_j\left(K+\frac{p}{\rho}\right)
-2\nu_{\mathrm{s}}u_iD_{ij}
-\frac{u_i\sigma_{ij}}{\rho},
\label{eq:supp_JK_full}\\
\varepsilon_{\mathrm{s}}={}&2\nu_{\mathrm{s}}D_{ij}D_{ij},
\qquad F=u_i f_i.
\label{eq:supp_eps_F_full}
\end{align}

Contracting equation~\eqref{eq:supp_B} with \(2K_pB_{ij}\) uses
\begin{equation}
B_{ij}a_{i\ell}B_{\ell j}=0,
\qquad
B_{ij}B_{i\ell}G_{\ell j}=C_{ij}G_{ij},
\qquad
B_{ij}(B^{-1})_{ij}=d,
\label{eq:supp_B_contractions}
\end{equation}
and gives
\begin{equation}
\partial_tP+\partial_{x_k}(u_kP)
=+\phi-\frac{P}{\lambda}.
\label{eq:supp_P_unfiltered}
\end{equation}
The second equality in equation~\eqref{eq:supp_K_P_phi} follows because
\(C_{ij}\) and \(\sigma_{ij}\) are symmetric, while
\(\delta_{ij}G_{ij}=\partial_{x_i}u_i=0\). Thus \(\phi\) converts kinetic
and polymer trace energy with opposite signs.
The square-root formulation does not alter this one-point equation; it makes
the trace energy quadratic in \(B_{ij}\), permitting an exact resolved
and subfilter energy split, as used in the spectral analysis of
\citet{Nguyen2016}.

Filtering these quantities gives the total filtered energies
\begin{equation}
[K]^{(r)}=\frac{1}{2}[u_i u_i]^{(r)},
\qquad
[P]^{(r)}=K_p([B_{ij}B_{ij}]^{(r)}-d).
\label{eq:supp_filtered_totals}
\end{equation}
We define the resolved large-scale reservoirs
\begin{equation}
K^{(r)}=\frac{1}{2}u_i^{(r)}u_i^{(r)},
\qquad
P^{(r)}=K_p(B_{ij}^{(r)}B_{ij}^{(r)}-d),
\label{eq:supp_resolved_reservoirs}
\end{equation}
and the subfilter reservoirs
\begin{equation}
K_{\mathrm{SFS}}^{(r)}=\frac{1}{2}
\left([u_i u_i]^{(r)}-u_i^{(r)}u_i^{(r)}\right),
\qquad
P_{\mathrm{SFS}}^{(r)}=K_p\left(
[B_{ij}B_{ij}]^{(r)}
-B_{ij}^{(r)}B_{ij}^{(r)}\right).
\label{eq:supp_sgs_reservoirs}
\end{equation}
For a normalised non-negative filter kernel, \(K_{\mathrm{SFS}}^{(r)}\geq0\)
and \(P_{\mathrm{SFS}}^{(r)}\geq0\), and
\begin{equation}
[K]^{(r)}=K^{(r)}+K_{\mathrm{SFS}}^{(r)},
\qquad
[P]^{(r)}=P^{(r)}+P_{\mathrm{SFS}}^{(r)}.
\label{eq:supp_reservoir_sum}
\end{equation}

\subsection*{Resolved and subfilter kinetic-energy equations}

Filtering the momentum equation gives
\begin{equation}
\begin{split}
\partial_tu_i^{(r)}
+u_j^{(r)}\partial_{x_j}u_i^{(r)}
={}&-\partial_{x_j}\tau^u_{ij}
-\frac{1}{\rho}\partial_{x_i}p^{(r)}
+2\nu_{\mathrm{s}}\partial_{x_j}D_{ij}^{(r)}
\\
&+\frac{1}{\rho}\partial_{x_j}\sigma_{ij}^{(r)}
+f_i^{(r)},
\end{split}
\label{eq:supp_momentum_filtered}
\end{equation}
where
\begin{equation}
\tau^u_{ij}=[u_i u_j]^{(r)}-u_i^{(r)}u_j^{(r)}.
\label{eq:supp_tau_u}
\end{equation}
Contracting equation~\eqref{eq:supp_momentum_filtered} with \(u_i^{(r)}\)
and applying the product rule yields
\begin{equation}
\begin{split}
\partial_tK^{(r)}+\partial_{x_j}J^{(r)}_{K,j}
={}&\tau^u_{ij}G_{ij}^{(r)}
-\frac{\sigma_{ij}^{(r)}G_{ij}^{(r)}}{\rho}
-2\nu_{\mathrm{s}}D_{ij}^{(r)}D_{ij}^{(r)}
+u_i^{(r)}f_i^{(r)},
\end{split}
\label{eq:supp_K_resolved_expanded}
\end{equation}
where
\begin{equation}
J^{(r)}_{K,j}=
u_j^{(r)}\left(K^{(r)}+\frac{p^{(r)}}{\rho}\right)
+u_i^{(r)}\tau^u_{ij}
-2\nu_{\mathrm{s}}u_i^{(r)}D_{ij}^{(r)}
-\frac{u_i^{(r)}\sigma_{ij}^{(r)}}{\rho}.
\label{eq:supp_JK_resolved}
\end{equation}
The terms in equation~\eqref{eq:supp_K_resolved_expanded} are written as
\begin{subequations}\label{eq:supp_kinetic_work_terms}
\begin{align}
\Pi_K^{(r)}={}&-\tau^u_{ij}G_{ij}^{(r)},
\label{eq:supp_tau_PiK}\\
\mathcal W_L^{(r)}={}&
\frac{\sigma_{ij}^{(r)}G_{ij}^{(r)}}{\rho}
=2K_p[B_{i\ell}B_{\ell j}]^{(r)}G_{ij}^{(r)},
\label{eq:supp_WL}\\
\varepsilon_{\mathrm{s}}^{(r)}={}&
2\nu_{\mathrm{s}}D_{ij}^{(r)}D_{ij}^{(r)},
\qquad
F^{(r)}=u_i^{(r)}f_i^{(r)}.
\label{eq:supp_eps_F_resolved}
\end{align}
\end{subequations}
The resolved kinetic-energy equation therefore becomes
\begin{equation}
\partial_tK^{(r)}+\partial_{x_j}J^{(r)}_{K,j}
=-\Pi_K^{(r)}-\mathcal W_L^{(r)}
-\varepsilon_{\mathrm{s}}^{(r)}+F^{(r)}.
\label{eq:supp_K_resolved}
\end{equation}
Here \(\Pi_K^{(r)}\) transfers kinetic energy between the retained and
subfilter scales, while \(\mathcal W_L^{(r)}\) represents kinetic--polymer
energy conversion within the retained motions. Positive \(\Pi_K^{(r)}\)
denotes transfer from \(K^{(r)}\) to \(K_{\mathrm{SFS}}^{(r)}\), and positive
\(\mathcal W_L^{(r)}\) denotes conversion from retained kinetic energy to
polymer energy. The remaining terms describe viscous dissipation and forcing
at the retained scales.

Filtering equation~\eqref{eq:supp_K_unfiltered} and subtracting
equation~\eqref{eq:supp_K_resolved} first gives
\begin{equation}
\begin{split}
\partial_tK_{\mathrm{SFS}}^{(r)}
+\partial_{x_j}J_{K,\mathrm{SFS},j}^{(r)}
={}&\Pi_K^{(r)}
-\left([\phi]^{(r)}
-\frac{\sigma_{ij}^{(r)}G_{ij}^{(r)}}{\rho}\right)
\\
&-\left([\varepsilon_{\mathrm{s}}]^{(r)}
-\varepsilon_{\mathrm{s}}^{(r)}\right)
+\left([F]^{(r)}-F^{(r)}\right).
\end{split}
\label{eq:supp_K_sgs_expanded}
\end{equation}
The remaining terms are collected as
\begin{subequations}\label{eq:supp_kinetic_sfs_terms}
\begin{align}
J_{K,\mathrm{SFS},j}^{(r)}={}&
[J_{K,j}]^{(r)}-J^{(r)}_{K,j},
\label{eq:supp_JK_sgs}\\
\mathcal W_T^{(r)}={}&[\phi]^{(r)}
=2K_p[B_{i\ell}B_{\ell j}G_{ij}]^{(r)},
\label{eq:supp_WT}\\
\mathcal W_S^{(r)}={}&\mathcal W_T^{(r)}-\mathcal W_L^{(r)}
=2K_p\left(
[B_{i\ell}B_{\ell j}G_{ij}]^{(r)}
-[B_{i\ell}B_{\ell j}]^{(r)}G_{ij}^{(r)}
\right),
\label{eq:supp_WS}\\
\varepsilon_{\mathrm{s},\mathrm{SFS}}^{(r)}={}&
[\varepsilon_{\mathrm{s}}]^{(r)}-\varepsilon_{\mathrm{s}}^{(r)}
=2\nu_{\mathrm{s}}\left(
[D_{ij}D_{ij}]^{(r)}-D_{ij}^{(r)}D_{ij}^{(r)}\right),
\label{eq:supp_eps_sgs}\\
F_{\mathrm{SFS}}^{(r)}={}&
[F]^{(r)}-F^{(r)}
=[u_i f_i]^{(r)}-u_i^{(r)}f_i^{(r)}.
\label{eq:supp_F_sgs}
\end{align}
\end{subequations}
The subfilter kinetic-energy equation is thus
\begin{equation}
\partial_tK_{\mathrm{SFS}}^{(r)}
+\partial_{x_j}J_{K,\mathrm{SFS},j}^{(r)}
=+\Pi_K^{(r)}-\mathcal W_S^{(r)}
-\varepsilon_{\mathrm{s},\mathrm{SFS}}^{(r)}
+F_{\mathrm{SFS}}^{(r)}.
\label{eq:supp_K_sgs}
\end{equation}
Here \(\mathcal W_S^{(r)}\) is the corresponding kinetic--polymer conversion
at the subfilter scales: positive values denote conversion from subfilter
kinetic energy to polymer energy, whereas negative values denote the reverse.
The terms \(\varepsilon_{\mathrm{s},\mathrm{SFS}}^{(r)}\) and
\(F_{\mathrm{SFS}}^{(r)}\) represent subfilter viscous dissipation and forcing,
respectively.

\subsection*{Resolved square-root polymer-energy equation}

Introduce the subfilter transport of \(B_{ij}\),
\begin{equation}
\tau^B_{kij}
=[u_kB_{ij}]^{(r)}-u_k^{(r)}B_{ij}^{(r)}.
\label{eq:supp_tauB}
\end{equation}
Filtering equation~\eqref{eq:supp_B} gives
\begin{align}
\partial_tB_{ij}^{(r)}
+u_k^{(r)}\partial_{x_k}B_{ij}^{(r)}
={}&-\partial_{x_k}\tau^B_{kij}
+[B_{i\ell}G_{\ell j}]^{(r)}
+[a_{i\ell}B_{\ell j}]^{(r)}
\notag\\
&+\frac{1}{2\lambda}
\left([(B^{-1})_{ij}]^{(r)}-B_{ij}^{(r)}\right).
\label{eq:supp_B_filtered}
\end{align}
Contracting equation~\eqref{eq:supp_B_filtered} with
\(2K_pB_{ij}^{(r)}\) and applying the product rule gives the complete
resolved polymer-energy equation
\begin{equation}
\begin{split}
\partial_tP^{(r)}
+\partial_{x_k}\!\left(
u_k^{(r)}P^{(r)}+2K_pB_{ij}^{(r)}\tau^B_{kij}\right)
={}&2K_p\tau^B_{kij}\partial_{x_k}B_{ij}^{(r)}
+2K_pB_{ij}^{(r)}[B_{i\ell}G_{\ell j}]^{(r)}
\\
&+2K_pB_{ij}^{(r)}[a_{i\ell}B_{\ell j}]^{(r)}
-\frac{K_p}{\lambda}\left(
B_{ij}^{(r)}B_{ij}^{(r)}
-B_{ij}^{(r)}[(B^{-1})_{ij}]^{(r)}\right).
\end{split}
\label{eq:supp_P_resolved_expanded}
\end{equation}
We define
\begin{subequations}\label{eq:supp_B_terms}
\begin{align}
J^{(r)}_{P,k}={}&
u_k^{(r)}P^{(r)}+2K_pB_{ij}^{(r)}\tau^B_{kij},
\label{eq:supp_JP_resolved}\\
\Pi_{P,\mathrm{adv}}^{(r)}={}&
-2K_p\tau^B_{kij}\partial_{x_k}B_{ij}^{(r)},
\label{eq:supp_PiP_adv}\\
\Pi_{\mathrm{ori}}^{(r)}={}&
-2K_pB_{ij}^{(r)}
[a_{i\ell}B_{\ell j}]^{(r)},
\label{eq:supp_Pi_ori}\\
\Pi_P^{(r)}={}&
\Pi_{P,\mathrm{adv}}^{(r)}+\Pi_{\mathrm{ori}}^{(r)},
\label{eq:supp_PiP_self}\\
\mathcal{S}_{B,G}^{(r)}={}&
2K_pB_{ij}^{(r)}
[B_{i\ell}G_{\ell j}]^{(r)},
\label{eq:supp_SBG}\\
R^{(r)}={}&\frac{K_p}{\lambda}\left(
B_{ij}^{(r)}B_{ij}^{(r)}
-B_{ij}^{(r)}\,[(B^{-1})_{ij}]^{(r)}
\right).
\label{eq:supp_R_resolved}
\end{align}
\end{subequations}

The compact resolved polymer-energy equation is therefore
\begin{equation}
\partial_tP^{(r)}+\partial_{x_k}J^{(r)}_{P,k}
=-\Pi_P^{(r)}+\mathcal{S}_{B,G}^{(r)}-R^{(r)}.
\label{eq:supp_P_resolved_direct}
\end{equation}
Here \(\Pi_P^{(r)}\) is the inter-scale transfer within the polymer-energy
reservoir. Positive values denote transfer from \(P^{(r)}\) to
\(P_{\mathrm{SFS}}^{(r)}\), whereas negative values denote the reverse.
The term \(\mathcal S_{B,G}^{(r)}\) is the stretching input to the retained
polymer energy, and \(R^{(r)}\) is its relaxation contribution.

\begin{figure}[p]
\centering
\includegraphics[width=\textwidth]{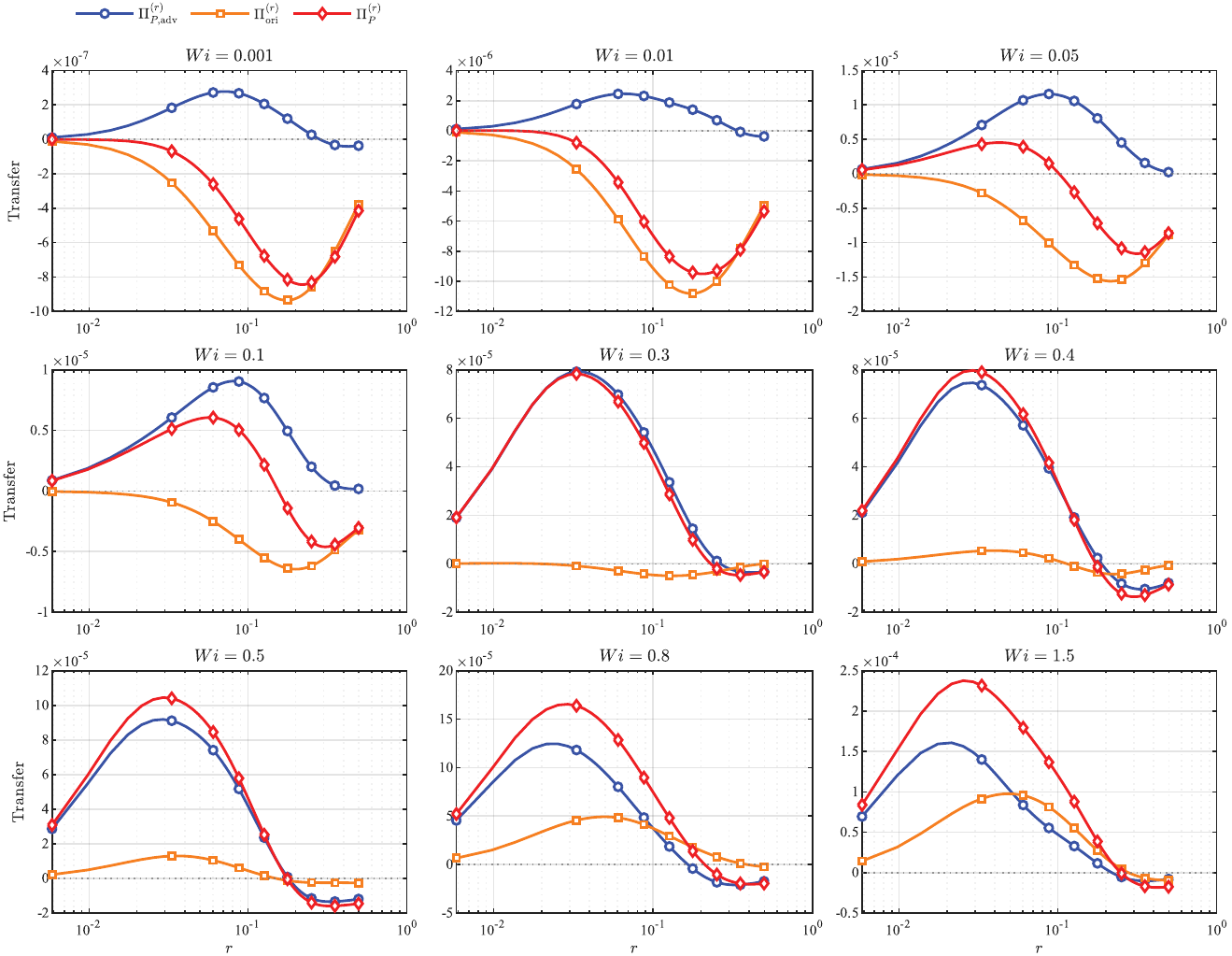}
\caption{Decomposition of the square-root polymer self-transfer for the nine
polymeric cases. The curves show the raw signed core-region means of
\(\Pi_{P,\mathrm{adv}}^{(r)}\), \(\Pi_{\mathrm{ori}}^{(r)}\), and
\(\Pi_P^{(r)}=\Pi_{P,\mathrm{adv}}^{(r)}+\Pi_{\mathrm{ori}}^{(r)}\) as
functions of Gaussian filter width \(r\); no peak normalisation is applied.
Positive transfer removes energy from the resolved polymer reservoir and
supplies its subfilter counterpart, while negative transfer denotes the
reverse direction. The ordinate multiplier is set independently in each
panel. Markers denote the evaluated filter widths and connecting curves are
guides to the eye.}
\label{fig:supp_PiP_decomposition}
\end{figure}

Figure~\ref{fig:supp_PiP_decomposition} shows the two contributions separately
for every polymeric case used in the transfer analysis. This decomposition is
useful because the orientation-compensation term can either reinforce or
offset the advective part; retaining only
\(\Pi_{P,\mathrm{adv}}^{(r)}\) would therefore not reproduce the
square-root polymer self-transfer.

\subsection*{Subfilter square-root polymer-energy equation}

Filtering equation~\eqref{eq:supp_P_unfiltered} gives
\begin{equation}
\partial_t[P]^{(r)}+\partial_{x_k}[u_kP]^{(r)}
=\mathcal W_T^{(r)}-\frac{[P]^{(r)}}{\lambda}.
\label{eq:supp_P_filtered_total}
\end{equation}
Subtracting equation~\eqref{eq:supp_P_resolved_direct} yields
\begin{equation}
\begin{split}
\partial_tP_{\mathrm{SFS}}^{(r)}
+\partial_{x_k}J_{P,\mathrm{SFS},k}^{(r)}
={}&-2K_p\tau^B_{kij}\partial_{x_k}B_{ij}^{(r)}
+2K_p\left(
[B_{i\ell}B_{\ell j}G_{ij}]^{(r)}
-B_{ij}^{(r)}[B_{i\ell}G_{\ell j}]^{(r)}
\right)
\\
&-2K_pB_{ij}^{(r)}[a_{i\ell}B_{\ell j}]^{(r)}
-\frac{[P]^{(r)}}{\lambda}
+\frac{K_p}{\lambda}\left(
B_{ij}^{(r)}B_{ij}^{(r)}
-B_{ij}^{(r)}[(B^{-1})_{ij}]^{(r)}
\right).
\end{split}
\label{eq:supp_P_sgs_expanded}
\end{equation}
We define
\begin{subequations}\label{eq:supp_P_sgs_terms}
\begin{align}
J_{P,\mathrm{SFS},k}^{(r)}={}&[u_kP]^{(r)}-J^{(r)}_{P,k}
\notag\\
={}&u_k^{(r)}P_{\mathrm{SFS}}^{(r)}
+\left([u_kP]^{(r)}-u_k^{(r)}[P]^{(r)}\right)
-2K_pB_{ij}^{(r)}\tau^B_{kij},
\label{eq:supp_JP_sgs}\\
R_{\mathrm{SFS}}^{(r)}={}&\frac{[P]^{(r)}}{\lambda}-R^{(r)}.
\label{eq:supp_R_sgs}
\end{align}
\end{subequations}
The compact subfilter polymer-energy equation is therefore
\begin{equation}
\partial_tP_{\mathrm{SFS}}^{(r)}
+\partial_{x_k}J_{P,\mathrm{SFS},k}^{(r)}
=+\Pi_P^{(r)}
+\left(\mathcal W_T^{(r)}-\mathcal S_{B,G}^{(r)}\right)
-R_{\mathrm{SFS}}^{(r)}.
\label{eq:supp_P_sgs}
\end{equation}
Here positive \(\Pi_P^{(r)}\) supplies the subfilter polymer reservoir from
its retained counterpart. The term
\(\mathcal W_T^{(r)}-\mathcal S_{B,G}^{(r)}\) is the stretching input to
\(P_{\mathrm{SFS}}^{(r)}\), and \(R_{\mathrm{SFS}}^{(r)}\) is its relaxation
contribution.

\subsection*{Direct four-reservoir system}

Collecting the directly derived terms gives
\begin{subequations}\label{eq:supp_four_budget_direct}
\begin{align}
\partial_tK^{(r)}
+\partial_{x_j}J_{K,j}^{(r)}
={}&-\Pi_K^{(r)}-\mathcal W_L^{(r)}
-\varepsilon_{\mathrm{s}}^{(r)}+F^{(r)},
\label{eq:supp_four_K_res}\\
\partial_tK_{\mathrm{SFS}}^{(r)}
+\partial_{x_j}J_{K,\mathrm{SFS},j}^{(r)}
={}&+\Pi_K^{(r)}-\mathcal W_S^{(r)}
-\varepsilon_{\mathrm{s},\mathrm{SFS}}^{(r)}+F_{\mathrm{SFS}}^{(r)},
\label{eq:supp_four_K_sgs}\\
\partial_tP^{(r)}
+\partial_{x_j}J_{P,j}^{(r)}
={}&-\Pi_P^{(r)}+\mathcal S_{B,G}^{(r)}-R^{(r)},
\label{eq:supp_four_P_res}\\
\partial_tP_{\mathrm{SFS}}^{(r)}
+\partial_{x_j}J_{P,\mathrm{SFS},j}^{(r)}
={}&+\Pi_P^{(r)}
+\left(\mathcal W_T^{(r)}-\mathcal S_{B,G}^{(r)}\right)
-R_{\mathrm{SFS}}^{(r)}.
\label{eq:supp_four_P_sgs}
\end{align}
\end{subequations}
The divergences \(\partial_{x_j}J_{q,j}\) are physical-space transport terms.
The pairs \(\mp\Pi_K^{(r)}\) and \(\mp\Pi_P^{(r)}\) conservatively transfer
kinetic and polymer energy, respectively, between the retained and subfilter
scales. Since
\(\mathcal W_T^{(r)}=\mathcal W_L^{(r)}+\mathcal W_S^{(r)}
=[\phi]^{(r)}\), all internal inter-scale transfer and kinetic--polymer
conversion terms cancel when the four equations are added, yielding
\begin{equation}
\partial_t([K]^{(r)}+[P]^{(r)})
+\partial_{x_j}\left([J_{K,j}]^{(r)}
+[u_jP]^{(r)}\right)
=-[\varepsilon_{\mathrm{s}}]^{(r)}
-\frac{[P]^{(r)}}{\lambda}+[F]^{(r)}.
\label{eq:supp_total_closure}
\end{equation}

\subsection*{Fourier--Lin budgets and kinetic--polymer exchange}

Following the spectral formulation of \citet{Nguyen2016}, Fourier transforming
the momentum and square-root equations and summing the modal budgets over each
shell gives
\begin{subequations}\label{eq:supp_Lin_equations}
\begin{align}
\partial_tE_K(\kappa)={}&T_K(\kappa)-D_K(\kappa)
+S_{P\rightarrow K}(\kappa)+F_K(\kappa),
\label{eq:supp_Lin_K}\\
\partial_tE_P(\kappa)={}&T_P(\kappa)+T_a(\kappa)
+S_{K\rightarrow P}(\kappa)-D_P(\kappa).
\label{eq:supp_Lin_P}
\end{align}
\end{subequations}
Here \(T_K\) and \(T_P+T_a\) are the nonlinear inter-scale redistribution
terms, \(D_K\) and \(D_P\) are the solvent-dissipation and polymer-relaxation
spectra, and \(F_K\) is the forcing spectrum. The directional kinetic--polymer
exchange terms required here are
\begin{subequations}\label{eq:supp_directional_exchange}
\begin{align}
S_{P\rightarrow K}(\kappa)={}&\frac{1}{\rho}
\sum_{\boldsymbol{k}\in\mathcal K_{\kappa}}
\operatorname{Re}\left[
\widehat u_i^*(\boldsymbol{k})
\,\mathrm{i}k_j\widehat\sigma_{ij}(\boldsymbol{k})\right],
\label{eq:supp_SPK_shell}\\
S_{K\rightarrow P}(\kappa)={}&2K_p
\sum_{\boldsymbol{k}\in\mathcal K_{\kappa}}
\operatorname{Re}\left[
\widehat B_{ij}^*(\boldsymbol{k})
\widehat{B_{i\ell}G_{\ell j}}(\boldsymbol{k})\right].
\label{eq:supp_SKP_shell}
\end{align}
\end{subequations}
Positive \(S_{P\rightarrow K}\) denotes polymer-to-kinetic input to the
receiving kinetic shell, whereas positive \(S_{K\rightarrow P}\) denotes
kinetic-to-polymer input to the receiving polymer shell.

The signs and the global cancellation follow directly from Parseval's identity
and periodic integration by parts. Specifically,
\begin{subequations}\label{eq:supp_exchange_spectral_sum}
\begin{align}
\sum_{\kappa}S_{P\rightarrow K}(\kappa)
&=\frac{1}{\rho}\left\langle
u_i\partial_{x_j}\sigma_{ij}\right\rangle
=-\frac{1}{\rho}\left\langle
\sigma_{ij}\partial_{x_j}u_i\right\rangle
=-\langle\phi\rangle,
\label{eq:supp_SPK_sum}\\
\sum_{\kappa}S_{K\rightarrow P}(\kappa)
&=2K_p\left\langle
B_{ij}B_{i\ell}G_{\ell j}\right\rangle
=2K_p\left\langle C_{ij}G_{ij}\right\rangle
=+\langle\phi\rangle.
\label{eq:supp_SKP_sum}
\end{align}
\end{subequations}
Here \(\sigma_{ij}=\sigma_{ji}\) removes the antisymmetric part of the
velocity gradient, and incompressibility gives
\(\delta_{ij}G_{ij}=0\).

\subsection*{Relation between the filtered and Fourier--Lin formulations}

The filtered and Fourier--Lin formulations describe the same nonlinear
dynamics in complementary scale representations. FST gives the transfer across
a selected filter scale, while the Lin equations resolve the corresponding
redistribution among Fourier shells. For the Gaussian filter in
equation~\eqref{eq:supp_gaussian_filter}, the spatially averaged filter-space
transfers are related to the corresponding Lin transfer spectra by the
full-periodic-domain identities
\begin{subequations}\label{eq:supp_gaussian_flux_correspondence}
\begin{align}
\left\langle\Pi_K^{(r)}\right\rangle
={}&-\int_0^\infty
\left|\widehat{\mathcal G}^{(r)}(\kappa)\right|^2
T_K(\kappa)\,\dd\kappa,
\label{eq:supp_gaussian_PiK_correspondence}\\
\left\langle\Pi_P^{(r)}\right\rangle
={}&-\int_0^\infty
\left|\widehat{\mathcal G}^{(r)}(\kappa)\right|^2
\left[T_P(\kappa)+T_a(\kappa)\right],\dd\kappa.
\label{eq:supp_gaussian_PiP_correspondence}
\end{align}
\end{subequations}
Here
\(\lvert\widehat{\mathcal G}^{(r)}(\kappa)\rvert^2
=\exp[-r^2\kappa^2/(2\pi^2)]\). The squared multiplier occurs because both the
receiving field and its nonlinear source are filtered in the physical-space
inner product, and it smoothly suppresses progressively higher wavenumbers. On
the discrete periodic grid, the integrals in
equation~\eqref{eq:supp_gaussian_flux_correspondence} are evaluated by the
corresponding Fourier sums.

The periodic test uses prescribed smooth fields rather than a dynamically
evolved Oldroyd--B solution. On the square
\((x_1,x_2)\in[0,2\pi)^2\), the velocity is generated from the streamfunction
\begin{subequations}\label{eq:supp_periodic_test_fields}
\begin{align}
\psi={}&\sin x_1\sin x_2+0.17\sin(2x_1-x_2)
+0.11\cos(x_1+3x_2)+0.09\cos(3x_1+0.37),
\label{eq:supp_periodic_test_psi}\\
\boldsymbol{u}={}&
(\partial_{x_2}\psi,-\partial_{x_1}\psi).
\label{eq:supp_periodic_test_u}
\end{align}
The square-root conformation field is prescribed as
\begin{align}
\boldsymbol B={}&\boldsymbol R^{\mathsf T}(\theta)
\begin{pmatrix}b_1&0\\0&b_2\end{pmatrix}
\boldsymbol R(\theta),
\qquad
\boldsymbol R(\theta)=
\begin{pmatrix}\cos\theta&-\sin\theta\\
\sin\theta&\cos\theta\end{pmatrix},
\label{eq:supp_periodic_test_B}\\
\theta={}&0.35\sin(x_1-x_2),\qquad
b_1=1.30+0.16\sin(2x_1+x_2),\qquad
b_2=1.08+0.10\cos(x_1-2x_2).
\label{eq:supp_periodic_test_parameters}
\end{align}
\end{subequations}
This construction gives a divergence-free velocity and a symmetric
positive-definite \(\boldsymbol B\), with
\(\boldsymbol C=\boldsymbol B^2\). The test sets \(\rho=1\) and
\(K_p=0.37\), and evaluates all derivatives and filters spectrally on the
\(96\times96\) periodic grid.

\begin{figure}[p]
\centering
\includegraphics[width=\textwidth]{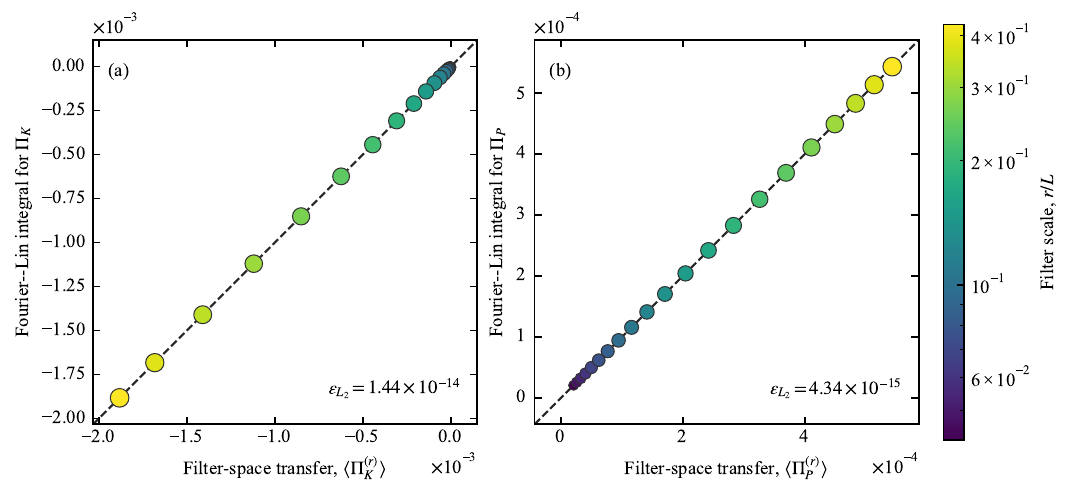}
\caption{Direct comparison of filter-space transfers and Gaussian-weighted
Fourier--Lin integrals for the prescribed periodic fields: (a) kinetic-energy
and (b) polymer-energy transfer. Each marker represents one of 21 filter
scales; colour and marker size indicate \(r/L\), and dashed lines denote
equality.}
\label{fig:supp_fst_fourier_periodic}
\end{figure}

The correspondence was then tested on these prescribed fields at 21 Gaussian
filter scales over \(0.042\leq r/L\leq0.424\), as shown in
figure~\ref{fig:supp_fst_fourier_periodic}. The Lin transfer spectra are
weighted by \(\lvert\widehat{\mathcal G}^{(r)}\rvert^2\) at each scale, as in
equation~\eqref{eq:supp_gaussian_flux_correspondence}. At each scale, the
filter-space value is evaluated directly from filtered physical-space
products, whereas the Fourier value is assembled independently from the
weighted Lin transfer spectra. The relative \(L_2\) differences are
\(1.44\times10^{-14}\) for \(\Pi_K\) and \(4.34\times10^{-15}\) for
\(\Pi_P\), verifying the integral relation across the tested scales.

\section*{Additional core-region statistics}

\subsection*{Pseudo-enstrophy-transfer increments}

\begin{figure}[p]
\centering
\includegraphics[width=\textwidth]{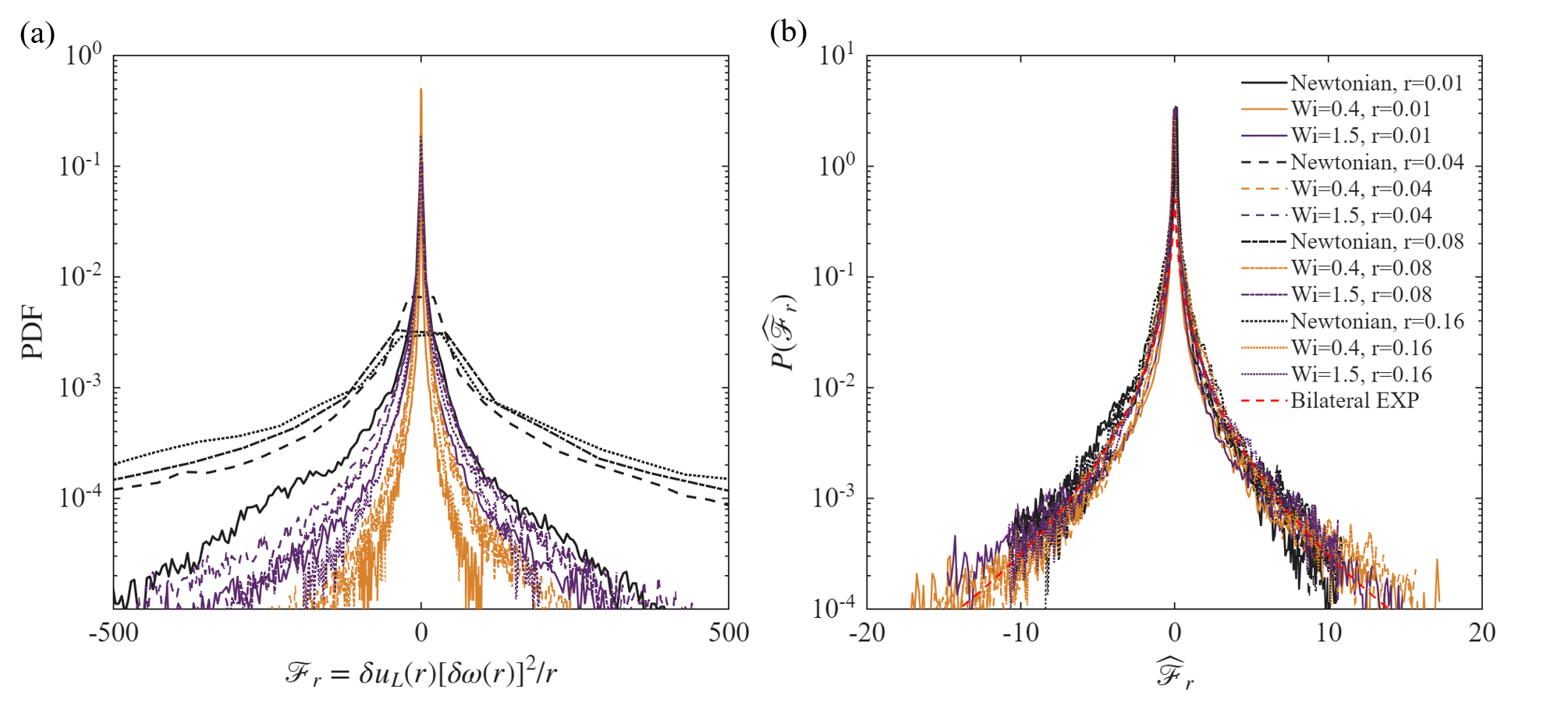}
\caption{Core-region PDFs of the pseudo-enstrophy-transfer increment
\(\mathcal F_r\) for the Newtonian, \(Wi=0.4\) and \(Wi=1.5\) cases at
\(r=0.01\), 0.04, 0.08 and 0.16. (a) Raw PDFs. (b) Centred and RMS-normalised
PDFs; the red dashed curve is a two-sided stretched-exponential reference.}
\label{fig:supp_enstrophy_increment}
\end{figure}

We additionally consider the local increment quantity
\begin{equation}
\mathcal F_r=\frac{\delta u_L(r)[\delta\omega(r)]^2}{r},
\label{eq:supp_enstrophy_increment}
\end{equation}
where the longitudinal velocity and vorticity increments from the two
coordinate directions are pooled. The raw PDFs in
figure~\ref{fig:supp_enstrophy_increment}a become markedly narrower after
polymer addition, indicating weaker fluctuations and fewer large-amplitude
pseudo-enstrophy-transfer events over the sampled separations. This suppression
is strongest for the representative \(Wi=0.4\) case; the distributions broaden
again at \(Wi=1.5\), but remain narrower than their Newtonian counterparts.
The polymer-induced reduction is therefore non-monotonic between the two
polymeric states shown, consistent with the two-regime behaviour of the
transfer statistics reported in the main text. After centring and RMS
normalisation, the PDFs collapse approximately onto a common two-sided
stretched-exponential reference
(figure~\ref{fig:supp_enstrophy_increment}b). Polymer addition thus primarily
changes the fluctuation amplitude, whereas the normalised form of the
intermittent events remains broadly similar. This stretched-exponential form is
consistent with experimental enstrophy-transfer statistics in Newtonian and
polymer-laden 2D turbulence \citep{Kellay2000,Kellay2004}.

\subsection*{Local Weissenberg number}

\begin{figure}[p]
\centering
\includegraphics[width=\textwidth]{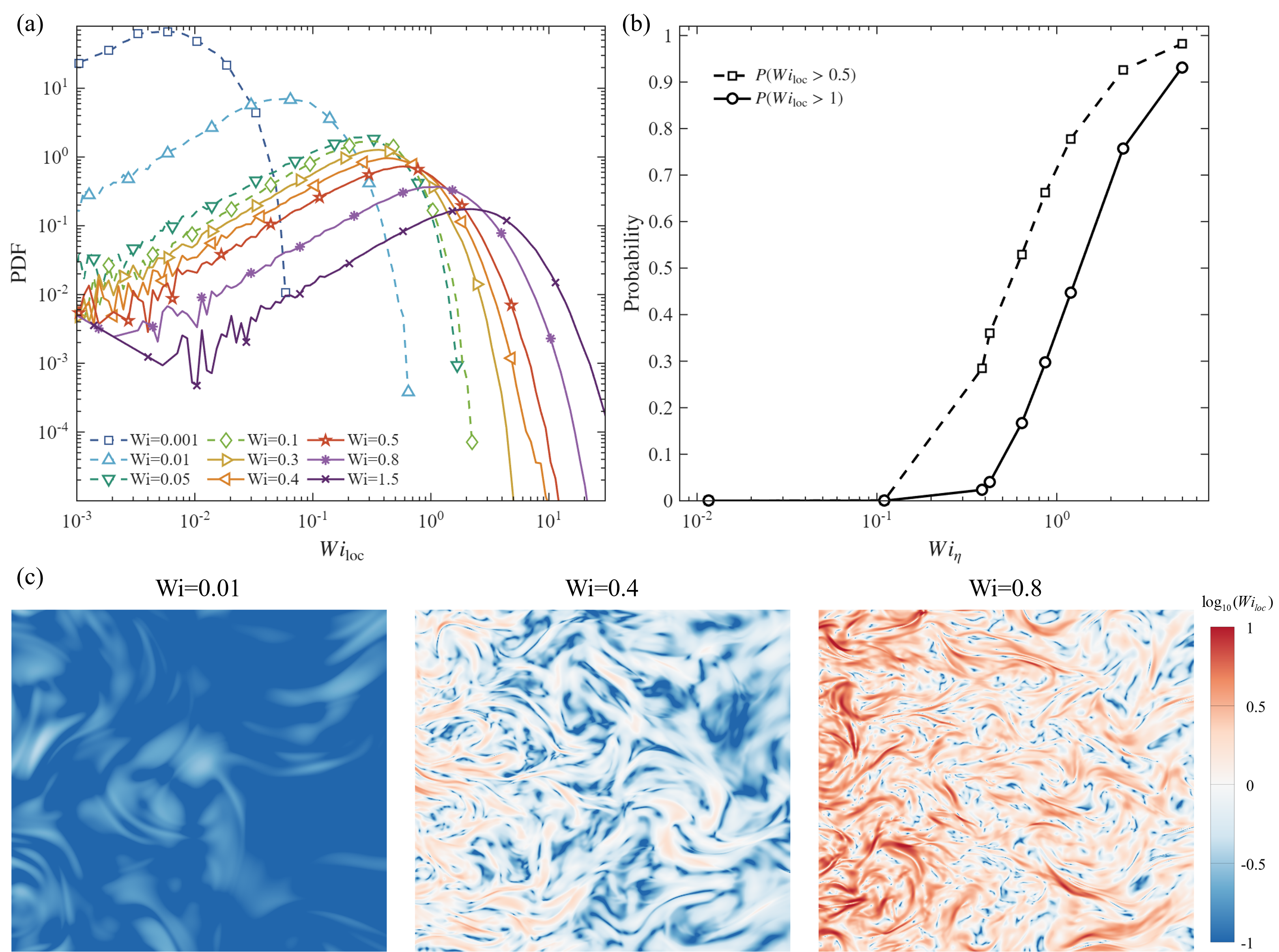}
\caption{Local-Weissenberg-number diagnostics in the core region: (a) PDFs
of \(Wi_{\mathrm{loc}}\); (b) fractions of sampled points satisfying
\(Wi_{\mathrm{loc}}>0.5\) and \(Wi_{\mathrm{loc}}>1\) as functions of
\(Wi_{\eta}\); (c) fields of \(\log_{10}(Wi_{\mathrm{loc}})\) at
\(Wi=0.01\), 0.4 and 0.8, from left to right.}
\label{fig:supp_Wiloc}
\end{figure}

The spatially heterogeneous elastic response is quantified by
\begin{equation}
Wi_{\mathrm{loc}}=\lambda\left(2D_{ij}D_{ij}\right)^{1/2}.
\label{eq:supp_Wiloc}
\end{equation}
The PDFs shift and broaden towards larger \(Wi_{\mathrm{loc}}\), while
\(P(Wi_{\mathrm{loc}}>0.5)\) and \(P(Wi_{\mathrm{loc}}>1)\), which measure the
fractions of sampled points above the two thresholds, rise rapidly across the
transition range (figure~\ref{fig:supp_Wiloc}a,b). The change therefore extends
beyond isolated extreme events: regions with \(Wi_{\mathrm{loc}}=O(1)\) occupy
an increasing fraction of the core region. The fields in
figure~\ref{fig:supp_Wiloc}c provide the spatial counterpart of these
statistics. Regions of elevated
\(Wi_{\mathrm{loc}}\) are sparse in the weak-elastic regime, spread along
elongated shear and inter-vortical regions through the transition, and form a
more extensive filamentary network in the elastic-activated regime. The
distributional change thus reflects the spatial expansion of regions with a
strong local elastic response rather than a spatially uniform increase. This
interpretation is consistent with the flow-topology changes reported in the
main text. The levels 0.5 and 1 are used only as diagnostic thresholds, rather
than strict coil--stretch criteria.

\clearpage
\bibliographystyle{latex_support/jfm}
\bibliography{latex_support/jfm}

\clearpage

